\documentclass[12pt]{article}
\usepackage{epsfig}
\usepackage{graphicx}

\def\beq{\begin{equation}}
\def\eeq#1{\label{#1}\end{equation}}
\def\eeqn{\end{equation}}

\def\beqa{\begin{eqnarray}}
\def\eeqa#1{\label{#1}\end{eqnarray}}
\def\eeqan{\end{eqnarray}}

\let\bar=\overbar

\def\Dslash{\not{\hbox{\kern-4pt $D$}}}
\def\dslash{\not{\hbox{\kern-2pt $\del$}}}

\def\msb{{\bar{\ssstyle M \kern -1pt S}}}

\usepackage{fancyhdr,graphicx}
\def\Title#1{\begin{center} {\Large {\bf #1} } \end{center}}

\begin{document}

\Title{Pulsars: Macro-nuclei with 3-flavour symmetry}

\bigskip\bigskip


\begin{raggedright}

{\it
Renxin Xu\\
School of Physics and KIAA (Kavli Institute for Astronomy and Astrophysics),\\
Peking University, Beijing 100871, China}

\end{raggedright}

\begin{abstract}
A pulsar-like compact star is the rump left behind after a supernova where normal baryonic matter is intensely compressed by gravity, but the real state of such compressed baryonic matter is still not well understood because of the non-perturbative nature of the fundamental color interaction. We argue that pulsars could be of condensed matter of quark clusters, i.e., ``quark-cluster stars'' which distinguish from conventional neutron and quark stars. In comparison with 2-flavour symmetric micro-nuclei, a quark-cluster star could simply be considered as a macro-nucleus with 3-flavour symmetry. New research achievements both theoretical and observational are briefly presented.
\end{abstract}

\section{Could macro-nuclei be 3-flavour symmetrical?}

It is empirically known that stable nuclei tend to have equal numbers of protons ($Z$) and neutrons ($N$), although we still have not well understood the underlying physics up to now.
Generally, this observation are reflected in the so-called symmetry energy so that the mass formula of a nucleus with atomic number $A~(=Z+N)$ consists of a sum of five terms,
\begin{equation}
E(Z,N)=a_{\rm v}A+a_{\rm s}A^{2/3}+a_{\rm sym}{(N-Z)^2\over A}+a_{\rm c}{Z(Z-1)\over A^{1/3}}+a_{\rm p}\Delta(Z,N),
\end{equation}
with the first term of bulk energy, the second term of surface energy, the third term of {\it symmetry eneryg}, the fourth term of Coulomb energy, and the fifth term of pairing energy.
It is then evident that those nuclei with extremely unbalanced numbers of $Z$ and $N$ should be unstable if the third term dominates.

It is worth noting that the fact of $Z=N$ is equivalent to the 2-flavour ($u$ and $d$ flavours of quarks) symmetry in the quark model, and that nuclei with 2-flavour symmetry should be positively charged since the electric charge of $u$-quark is $+2/3$ but that of $d$-quark is $-1/3$.
Therefore, related leptons (i.e., electrons, $e$) have to participate in a system with 2-flavour symmetry in order to keep neutrality.
Nevertheless, in case of small number of quarks (to speak more strictly, quarks are localized in a scale not much larger than the Compton wavelength, $\lambda_{\rm c}=0.024$ \AA, so that electrons are non-relativistic), electrons would contribute negligible energy because $\alpha_{\rm em}\ll \alpha_{\rm s}$,\footnote{
Because of large $\alpha_{\rm s}$, quarks are localized in 2-flavour symmetrical nuclei ($N\simeq Z$) if the number of valence quarks is $\ll 10^9$. However, electrons are bound via electro-magnetic interaction with a relatively much smaller coupling constant, $\alpha_{\rm em}$.
From Heisenberg's uncertainty relation, one could infer the interaction energy between electrons and nucleus: $E\simeq e^2/l \simeq p^2/m_{\rm e}\sim \hbar^2/(m_{\rm e}l^2)$, with $l$ the length scale. One could then have,
$$
l \sim {\hbar^2/(m_{\rm e}e^2)} = {1\over \alpha_{\rm em}} {\hbar
c\over m_{\rm e}c^2}\sim {\rm \AA},\; \; E \sim {e^2\over
l} = \alpha_{\rm em}^2m_{\rm e}c^2\sim 10^{-5}~{\rm MeV}.
$$
It is also evident from the first equation above that the length scale of nuclei should be $\ll l$ due to the fact that $\alpha_{\rm s}\gg \alpha_{\rm em}$ and constituent quark mass $m_{\rm q}\simeq 300 {\rm MeV}\gg m_{\rm e}$ if the strong interaction could approximately be Coulomb-like.
Therefore, atoms would only produced in a universe in which $\alpha_{\rm s}\gg \alpha_{\rm em}$, and we are lucky enough to exist.
} %
where $\alpha_{\rm em}$ and $\alpha_{\rm s}$ are the coupling constants of electromagnetic and strong interactions, respectively. Note here that the nuclear energy scale is order of $\sim 10$ MeV, which is much higher than $m_{\rm e}c^2=0.511$ MeV, where $m_{\rm e}$ is the rest mass of electron.

How about 3-flavour ($u$, $d$ and $s$ flavours of quarks) symmetry in nature? Macro-nuclei\footnote{
Their baryon numbers are $\gg 10^9$, and thus quarks are localized in a scale $\gg \lambda_{\rm c}$ so that electrons are inside a nucleus and are relativistic.}
 could seemingly be 3-flavour symmetric for the three points provided below.
(1) Among the six flavours of quarks, three ($u$, $d$ and $s$) are light, with current masses $m_{\rm q}<10^2$ MeV, while others ($c$, $t$ and $b$) are heavy, with $m_{\rm q}>10^3$ MeV. It might be natural to expect a light-flavour symmetry (i.e., 3-flavour symmetry) restoration in a system.
(2) According to an order-of-magnitude estimation from either Heisenberg's relation (localized quarks) or Fermi energy (free quarks), the energy scale of dense matter at nuclear or supra-nuclear densities could certainly be $\sim 500$ MeV, being larger the mass difference ($\sim 100$ MeV) between $s$-quark and $u/d$-quarks. This means that strangeness could be easily excited in case of nuclear/supra-nuclear densities.
(3) No electron participates in a system if 3-flavour symmetry is fully restored, without broken, so that electron contributes really zero energy. However, for micro (scale of $\sim$ fm) nuclei where electron contributes also negligible energy, there could be 2-flavour (rather than 3-flavour) symmetry because $s$-quark mass ($\sim 10^2$ MeV) is larger than $u/d$-quark masses ($\sim 10$ MeV).
We now understand that it is more economical to have 2-flavour micro nuclei because of massive $s$-quark and negligible electron kinematic energy, whereas macro-/gigantic-nuclei might be 3-flavour symmetric.
We can also call dense matter with 3-flavour symmetry as strange quark-cluster matter, or simply {\em strange matter}, in which a number of quarks (6, 9, 12, and even 18) are grouped together in so-called quark clusters~\cite{Xu2003}.

How to create macro-nuclei (even gigantic) in the Universe?
Besides $\sim M_\odot$-mass strange stars produced via supernova, strange quark-cluster stars with masses as low as $\sim 10^{-2} M_\odot$ could result from accretion-induced collapse (AIC) of white dwarfs~\cite{Xu2005}; during both processes normal baryonic matter is intensely compressed by gravity.
Strange quark-nuggets and strange planets~\cite{Xu2006} around strange stars could be ejected during the birth of central compact star~\cite{xw2003}, or during collision of strange stars in a binary system spiral towards each other due to loss of orbital energy via gravitational waves~\cite{Madsen2005}.
Alternatively, quark-nuggets could also be produced during and survive in the hadronization of the early Universe~\cite{Witten1984}, and seed black holes could form by merger of these nuggets at redshift as high as $z\sim 6$~\cite{LX2010}.

In summary, though micro-nuclei in our dairy life are of 2-flavour ($u$ and $d$), macro-/gigantic-nuclei could be 3-flavour symmetrical (i.e., with strangeness), which could be created though astrophysical processes in various extreme cases.

\section{Observational consequences}

There could be some similarities and differences between a micro nucleus and a macro nucleus, listed as following.
\begin{itemize}
  \item {\em Similarity} 1: Both micro and macro nuclei are self-bound by the strong color interaction, in which quarks are localized in groups called generally as quark-clusters. We are sure that there are two kinds of quark-clusters inside micro nuclei, the proton (with structure {$uud$}) and neutron ({$udd$}), but don't know well the clusters in macro-nuclei due to the lack of detailed experiments related.

  \item {\em Similarity} 2: Since the strong interaction might not be very sensitive to flavour, the interaction between general quark-clusters should be similar to that of nucleon, which is found to be Lennard-Jones-like by both experiment and modeling. Especially, one could then expect a hard core~\cite{Wilczek2007} (or repulsive core) of the interaction potential though no direct experiment now hints this existence.

  \item {\em Difference} 1: The most crucial difference is the change of flavour degree of freedom, from two ($u$ and $d$) in micro nuclei to three ($u$, $d$ and $s$) in macro/gigantic nuclei. We could thus have following different aspects derived.

  \item {\em Difference} 2: The number of quarks in a quark-cluster is 3 for micro nuclei, but could be 6, 9, 12, and even 18 for macro nuclei, since the interaction between $\Lambda$-particles could be attractive~\cite{Beane2011,Inoue2011} so that no positive pressure can support a gravitational star of $\Lambda$-cluster matter. We therefore call proton/neutron as light quark-clusters, while the strange quark-cluster as heavy clusters because of (1) massive $s$-quark and (2) large number of quarks inside.

  \item {\em Difference} 3: A micro nucleus could be considered as a quantum system so that one could apply quantum mean-field theory, whereas the heavy clusters in strange matter may be classical particles since the quantum wavelength of massive clusters might be even smaller than the mean distance between them.

  \item {\em Difference} 4: The equation of state of strange matter would be stiffer (see \S 4.1 in \cite{LGX2013}) than that of nuclear matter because the clusters in former should be non-relativistic but relativistic in latter. The kinematic energy of a cluster in both micro and macro nuclei could be $\sim 0.5$ GeV, which is much smaller than the rest mass (generally $> 2$ GeV) of a strange quark-cluster.

  \item {\em Difference} 5: Condensed matter of strange quark-clusters could be in a solid state at low temperature much smaller than the interaction energy between clusters. We could then expect solid pulsars~\cite{Xu2003} in nature although an idea of solid nucleus was also addressed~\cite{Bertsch1974} previously.
\end{itemize}

Observation consequences could then be deduced from above points.
From {\em Similarity} 1, we may expect low-mass strange matter stars unavoidable and mass-radius relations significantly deviated from that of normal neutron stars, and also possible implications of surface (e.g., drifting sub-pulses of radio emission, non-atomic feature in thermal X-ray spectra, and super-Eddington radiative and clean fire ball of supernova, $\gamma$-ray bursts, and even anomalous X-ray pulsars (AXPs)/soft $\gamma$-ray repeaters (SGRs)). 4U 1746-37 could be a quark-cluster star with ultra low mass and small radius~\cite{Li2015}.
From {\em Difference} 1, as accretion rate increases, a corona, an atmosphere or a crust could form on the surface when normal non-strange matter accretes on to a strange quark-cluster star, which could be essential to understand Type I X-ray bursts, especially super-bursts~\cite{Xu2014}.
From {\em Similarity} 2 and {\em Difference} 4, we may expect massive pulsar-like compact stars, with masses as high as $3M_\odot$ and even higher~\cite{LX2009}. The hard core plays an important role for quark-cluster stars to stand against gravity, while degenerate pressure dominates in white dwarfs and conventional neutron stars.
From {\em Difference} 5, we provide alternative free energy (elastic and gravitational energies, rather than magnetic one) for AXPs/SGRs in the quark-cluster model, and two types of glitches detected would be naturally understood too~\cite{Zhou2014}.

The mass spectrum of strange quark-cluster star is very wide, from a few $M_\odot$, to $\sim 10^{-2} M_\odot$ and planet mass, and even to $\sim 10^{20}$ GeV $\sim 10^{-4}$ g of nugget. Their astrophysical manifestations are certainly different, which will be discussed here.
We now focus on the existence of non-strange matter above strange matter surface.
At the limit of low accretion, the micro nuclei accreted onto strange matter could have a temperature, $T$, approximating the gravitational energy,
\begin{equation}
kT\sim {GMm\over R},
\end{equation}
where $M\simeq \rho_{\rm av}(4\pi R^3)/3$, with $\rho_{\rm av}$ the average density and $R$ the radius, and $m=Am_{\rm u}$ is the mass of an accreted nucleus, with $A$ the mass number.
Due to the 3-flavour symmetry broken in strange matter, there exist a Coulomb barrier above strange matter surface, with a hight of $ZV_{\rm q}$, where $Z$ is the charge number of an accreted nucleus.
Significant barrier penetration would occur if $kT>\eta ZV_{\rm q}$, and therefore one could obtain the minimum mass for this process,
\begin{equation}
\begin{array}{lll}
M>M_{\rm min} & \simeq &\sqrt{3\over 4\pi}(Gm_{\rm u})^{-3/2}(Z/A)^{3/2}\eta^{3/2} V_{\rm q}^{3/2} \rho_{\rm av}^{-1/2}\\
& \sim & 1.8\times 10^{-4} (Z/A)^{3/2}\eta_{-4}^{3/2} V_{\rm q10}^{3/2} \rho_{\rm av2}^{-1/2} M_\odot,
\end{array}
\end{equation}
where $\rho_{\rm av}=2\rho_{\rm av2}\rho_0$ (nuclear density $\rho_0 = 2.8\times 10^{14}$ g/cm$^3$), $V_{\rm q}=(10~{\rm MeV})~V_{\rm q10}$, and $\eta=10^{-4}\eta_{-4}$.
Typical value of $M_{\rm min}$ is $\sim 60M_\oplus$, order of planet mass, but with radius of $\sim 500$ meters.
If cold strange objects with mass $<M_{\rm min}$ could form, the mass of normal non-strange matter accreted onto strange matter increases with time. Nevertheless, the mass may not increase significantly duo to the limit of the cosmic age, especially as single objects accreting from interstellar medium. A historical memory of the Universe could be stored there. Things are different for strange quark-nuggets, which could usually move relativistically (collision and fragmentation occurs there).
For strange object with mass $>M_{\rm min}$, an equilibrium between accretion of normal matter and penetration (converting to strange matter) could be settled.

\section{Conclusions}

Pulsars are proposed to be macro-nuclei with 3-flavour symmetry, and recent research achievements related are briefly summarized. We compare Micro- and macro-nuclei, showing similarities and differences between them.
In this paper, we are trying to focus on the researches of our pulsar group in order to see more comments and suggestions on the idea from the global colleagues professional and/or interested  in pulsar inner structure, but never trying to balance a presentation of the field.
We are very sorry for missing very important and interesting references which should be cited.

\subsection*{Acknowledgement}

We express our thanks to the organizers of the CSQCD IV conference for providing an excellent academic atmosphere which was the basis for inspiring discussions with all participants. The conference proceedings could make continuity of the science, from which we could benefit more.
This work is supported by the 973 program (No. 2012CB821800) and the NNSF (No. 11225314) of China.

\end{document}